\documentclass[aps,pra,eqsecnum,twocolumn]{revtex4}
\usepackage{graphicx}
\usepackage{bm}
\begin{document}
\preprint{}
\title{Angular-momentum nonclassicality by breaking classical
bounds on statistics}
\author{Alfredo Luis$^{1}$ and \'{A}ngel Rivas$^{2}$}
\affiliation{$1$ Departamento de \'{O}ptica, Facultad de Ciencias
F\'{\i}sicas, Universidad Complutense, 28040 Madrid, Spain \\
$^2$ Departamento de F\'{\i}sica Te\'orica I, Facultad
de Ciencias F\'{\i}sicas, Universidad Complutense, 28040 Madrid,
Spain}
\date{October 12, 2011}

\begin{abstract}
We derive simple practical procedures revealing the quantum behavior
of angular momentum variables by the violation of classical upper
bounds on the statistics. Data analysis is minimum and definite
conclusions are obtained without evaluation of moments, or any
other more sophisticated procedures. These nonclassical tests
are very general and independent of other typical quantum signatures
of nonclassical behavior such as sub-Poissonian statistics, squeezing,
or oscillatory statistics, being insensible to the nonclassical behavior
displayed by other variables.
\end{abstract}


\maketitle

\section{Introduction}

Nonclassicality is a key concept supporting the necessity of the
quantum theory \cite{MSZ,MH,ase,as,AT,vyt,SP,MM}. A customary
signature of nonclassical behavior is the failure of the
Glauber-Sudarshan $P$ phase-space representation to exhibit all
the properties of a classical probability density. This occurs
when $P$ takes negative values, or when it is more singular
than a delta function.

In a recent work we have derived exceedingly simple and robust
practical procedures to reveal the quantum nature of states and
measurements \cite{nos1,nos2}. These are upper bounds on the outcome
probabilities which are satisfied when the $P$ representative
is compatible with classical physics. The lack of compliance of
these statistical bounds is thus a nonclassical signature
so this provides sufficient, not necessary, criteria
of nonclassicality.

In this work we derive the classical upper bounds for
the statistics of angular momentum or spin components, this is
to say SU(2) variables. They are derived in terms of the
classical or nonclassical behavior of the SU(2) $P$ function
for states and measurements. This generalizes previous particular
examples considered in Ref. \cite{nos1}. For definiteness we
focus on quantum optics where SU(2) variables represent very
basic items such as polarization and two-beam interference.
The main properties of this approach are:

i) The violation of these bounds can be ascribed exclusively
to the nonclassical behavior of SU(2) variables, this is when
the SU(2) $P$ function takes negative values or is more singular
than a delta function, irrespective of the classical or nonclassical
behavior of other variables, such as light intensity (photon
number).

ii) We show that these SU(2) upper bounds are larger than the ones
derived from the quadrature $P$ function. In the bright limit they
coincide with the bounds for field quadratures.

iii) The only previously reported nonclassical spin property
is SU(2) squeezing \cite{su2s,k,BM,yk} (in passing we
explicitly demonstrate below that SU(2) squeezing is actually
an SU(2) nonclassical property). This approach generalizes and
simplifies the idea of SU(2) squeezing so that it can be easily
applied to any spin observable. This is achieved without involving
state reconstruction, i. e., without complete knowledge of the
SU(2) $P$ function or any other distribution \cite{GBB,opd}.

iv) Data analysis is reduced to minimum so that definite
conclusions can be obtained without evaluation of moments,
or any other more sophisticated data elaborations
\cite{MSZ,MH,ase,as,AT,vyt}. This is reflected on the
robustness under practical imperfections \cite{nos1,nos2}.

v) These nonclassical tests are in general independent of
other typical quantum signatures such as sub-Poissonian
statistics, squeezing, or oscillatory statistics \cite{MSZ}.
To show this we provide some examples of quantum states
violating classical bounds that present no such typical
nonclassical signatures.

In Sec. II we recall the main tools required to the quantum
description of angular-momentum variables, including SU(2)
squeezing and the classical upper bounds to the statistics
of arbitrary spin observables. In Sec. III we show that the
angular-momentum components are nonclassical observables.
We also derive the classical upper bounds for the statistics
of angular-momentum components, applying them to some relevant
states.

\section{SU(2) systems}

In this section we first recall basic material on SU(2)
states and observables relevant for the analysis of
their nonclassical properties. We also demonstrate that
SU(2) squeezing is actually an SU(2) nonclassical property.

\subsection{Angular momentum operators}

Arbitrary dimensionless angular momentum operators
$\bm{j} = (j_1, j_2 , j_3 )$ satisfy the commutation relations
\begin{equation}
\label{cr}
[ j_k ,j_\ell ] = i \sum_{n=1}^3 \epsilon_{k,\ell ,n}
j_n ,
\qquad [j_0,\bm{j} ] = \bm{0} ,
\end{equation}
where $\epsilon_{k,\ell,n}$ is the fully antisymmetric
tensor with $\epsilon_{1,2,3} =1$, and $j_0$ is defined
by the relation
\begin{equation}
\label{j2}
\bm{j}^2 = j_0 \left ( j_0 + 1 \right ) .
\end{equation}
\textit{Note that this implies that all quantities to be considered
throughout this work, including all plots,  are dimensionless.}

For the sake of completeness we take into account that
$j_0$ may be an operator. This is the case of two-mode
bosonic realizations where $j_0$ is proportional to the
number of particles. More specifically
\begin{eqnarray}
\label{So}
j_0 = \frac{1}{2} \left ( a^\dagger_1 a_1 + a^\dagger_2
a_2 \right ), & &
j_1 = \frac{1}{2} \left ( a^\dagger_2 a_1 + a^\dagger_1
a_2 \right ) , \nonumber \\ & & \\
j_2 = \frac{i}{2} \left ( a^\dagger_2 a_1 - a^\dagger_1
a_2 \right ),  & &
j_3 = \frac{1}{2} \left ( a^\dagger_1 a_1 - a^\dagger_2
a_2 \right ) , \nonumber
\end{eqnarray}
where $a_{1,2}$ are the annihilation operators of two
independent bosonic modes with $[a_j,a^\dagger_j ]=1$,
$[a_1, a_2 ] = [a_1, a^\dagger_2 ]=0$ \cite{SCH}.
We have the following correspondence
\begin{equation}
|j,m \rangle = |n_1 = j+m \rangle | n_2 = j-m \rangle ,
\end{equation}
between the $|j,m \rangle$ basis of simultaneous eigenvectors
of $j_3$ and $j_0$, with $j_3 |j,m \rangle = m |j,m \rangle$
and  $j_0 |j,m \rangle = j |j,m \rangle$, and the product of
two-mode number states $|n_1 \rangle | n_2 \rangle$, with
$a^\dagger_j a_j  |n_j \rangle = n_j | n_j \rangle$.
\textit{The quantum number $j$ represents the total number of
bosons. For most realistic and practical situations the number
of bosons usually rather large, so below we will consider
suitable approximations of results in the limit $j \gg 1$.}

Concerning physical realizations, $a_{1,2}$ can represent
the complex amplitude operators of two electromagnetic
field modes. The operators $\bm{j}$ describe the polarization
of transverse electromagnetic waves (representing the Stokes
operators) as well as two-beam interference. For material
systems $a_{1,2}$ can represent the annihilation operators
for two species of atoms in two different internal states,
for example. Angular momentum operators also serve to describe
the internal state of two-level atoms via the definitions
\begin{eqnarray}
j_0 = \frac{1}{2} \left ( |e \rangle \langle e | + |g \rangle
\langle g | \right ), & &
j_1 = \frac{1}{2} \left ( |g \rangle \langle e | + |e \rangle
\langle g | \right ) , \nonumber \\ & & \\
j_2 = \frac{i}{2} \left ( |g \rangle \langle e | - |e \rangle
\langle g | \right ),  & &
j_3 = \frac{1}{2} \left ( |e \rangle \langle e | - |g \rangle
\langle g | \right ) , \nonumber
\end{eqnarray}
where $|e,g \rangle$ are the excited and ground states.
This is formally an spin 1/2 where $j_{0,3}$ represent atomic
populations and $j_{1,2}$ the atomic dipole \cite{SZ}.
Collections of two-level atoms are described by composition
of the individual angular momenta. We recall that for
spin 1/2 spin nonclassicality is equivalent to entanglement
\cite{KCL}.

\subsection{Phase space representatives}

The SU(2) $Q$ and $P$ functions associated to an operator $A$
are defined after the SU(2) coherent states $|j, \Omega
\rangle $ \cite{ACGT}
\begin{widetext}
\begin{equation}
\label{cs}
|j, \Omega \rangle = \sum_{m = -j}^j \left ( \begin{array}{c}
2j \cr m+j \end{array} \right )^{1/2} \sin^{j-m}
\left ( \frac{\theta}{2} \right ) \cos^{j+m} \left (
\frac{\theta}{2} \right ) \exp [-i (j+m) \phi] |j,m \rangle ,
\end{equation}
\end{widetext}
with $\pi \geq \theta \geq 0$, and $\pi \geq \phi \geq - \pi$,
as
\begin{equation}
A = \int d^2 \Omega P(\Omega) | j, \Omega \rangle
\langle j, \Omega | , \quad Q(\Omega) = \frac{2j+1}{4 \pi}
\langle j , \Omega | A | j, \Omega \rangle,
\end{equation}
with $d^2 \Omega = \sin \theta d \theta d \phi$. They are suitably
normalized since
\begin{equation}
\label{nor}
\int d^2 \Omega P (\Omega) = \int d^2 \Omega Q(\Omega)
= \mathrm{tr} A .
\end{equation}

Arbitrary measurements are described by positive operator-valued
measures (POVMs) $\Delta_k$, such that the probability of the
outcome $k$ is $p_k = \mathrm{tr} ( \Delta_k \rho )$, where
$\rho$ is the measured state. In terms of the SU(2) phase-space
representatives the statistics can be expressed as
\begin{equation}
\label{pm}
p_k = \frac{4 \pi}{2j+1} \int d^2 \Omega P_k (\Omega)
Q (\Omega) = \frac{4 \pi}{2j+1} \int d^2 \Omega P (\Omega )
Q_k (\Omega)  ,
\end{equation}
where $P (\Omega)$ and $Q (\Omega)$ are the representatives of
the measured state $\rho$, while $P_k (\Omega)$ and $Q_k
(\Omega)$ are the ones associated to the POVM $\Delta_k$.

We say that the measurement is nonclassical when the $P$
representative of some $\Delta_k$ takes negative values or is
more singular than a delta function. In most practical situations
$\Delta_k$ define legitimate measuring states $\rho_k \propto
\Delta_k$ so that the measurement is nonclassical if and only
there is a nonclassical measuring state $\rho_k$.

\subsection{SU(2) squeezing}

This can be regarded as the first exclusively SU(2) nonclassical
property. In general terms, the idea of SU(2) squeezing means
reduced fluctuations below the level established by the SU(2)
coherent states \cite{ACGT}. There are several quantitative
implementations of this idea \cite{su2s,k,BM,yk}:

i) The less stringent squeezing criterion is
that the fluctuations of a $\bm{j}$ component $j_\perp$
orthogonal to the direction of $\langle \bm{j} \rangle$ (this
is that $\langle j_\perp \rangle =0$) must be lesser than in a
SU(2) coherent state, leading to
\begin{equation}
\label{sc}
(\Delta j_\perp )^2 < \frac{j}{2} .
\end{equation}

ii) SU(2) squeezing can be defined as equivalent to provide
larger interferometric resolution than coherent states, leading
to
\begin{equation}
\label{is}
\frac{\left ( \Delta j_\perp \right )^2}{\langle \bm{j}
\rangle^2} < \frac{1}{2j} .
\end{equation}
This implies the satisfaction of the most general squeezing
condition (\ref{sc}).

iii) Finally, there is also the idea of squeezing derived
from the uncertainty relations (focusing again on orthogonal
components)
\begin{equation}
\Delta j_{\perp, 1} \Delta j_{\perp, 2} \geq \frac{1}{2} |
\langle \bm{j} \rangle | ,
\end{equation}
so that SU(2) squeezing would mean
\begin{equation}
\label{urc}
(\Delta j_{\perp} )^2 < \frac{| \langle \bm{j} \rangle |}{2} ,
\end{equation}
which implies the satisfaction of both Eqs. (\ref{sc}) and
(\ref{is}). In particular, this is achieved by the SU(2)
intelligent states determined by the following eigenvalue
equation \cite{BM}
\begin{equation}
\label{eeis}
\left ( \eta j_{\perp,2} + i j_{\perp,1} \right ) | \psi
\rangle = 0 ,
\end{equation}
where $\eta$ is a real parameter. For $\eta =1$ they are SU(2)
coherent states so that uncertainty-relations squeezing (\ref{urc})
occurs for $\eta \neq 1$ and implies the satisfaction of the other
criteria (\ref{sc}) and (\ref{is}).

\subsubsection{SU(2) squeezing is an SU(2) nonclassical property}

Next we show that every SU(2) squeezed state has a nonclassical
SU(2) $P$ distribution. This completes the proof in Ref. \cite{yk}
where it was shown in bosonic realizations that SU(2) squeezing
implies nonclassical quadrature $P$ function.

To this end we focus on the most general criterion in Eq. (\ref{sc}).
Using the SU(2) $P$ representation we have
\begin{equation}
(\Delta j_\perp )^2 = \langle j_\perp^2 \rangle = \int d^2 \Omega
P (\Omega ) \langle j, \Omega |  j_\perp^2 | j, \Omega \rangle .
\end{equation}
It can be easily seen using SU(2) invariance that for any component
$j_u = \bm{u} \cdot \bm{j}$ with $\bm{u}^2 = 1$ we have the identity
\begin{equation}
\label{tr}
\langle j, \Omega | j_u^2 | j, \Omega \rangle = \frac{j}{2} +
\frac{2j-1}{2j} \langle j, \Omega | j_u | j , \Omega
\rangle ^2 .
\end{equation}
\textit{To demonstrate this relation we use SU(2) invariance (every SU(2)
coherent state can be obtained by applying an SU(2) transformation
to $| j, m=j \rangle$) so that}
\begin{equation}
\langle j, \Omega | j^k_u | j, \Omega \rangle =
\langle j, m=j | j^k_v | j, m=j \rangle
\end{equation}
\textit{where $| j, m=j \rangle$ is in the $j_0, j_3$ basis and
$\bm{v}$ is a unit real vector related with $\bm{u}$ by a rotation.
Using the bosonic representation (\ref{So}) the state $| j, m=j
\rangle$ becomes the photon number state $|n \rangle | 0 \rangle$
so that $\langle j, m=j | j_v | j, m=j \rangle = v_3 n/2$
and}
\begin{eqnarray}
\langle j, m=j | j_v^2 | j, m=j \rangle &=& (v_1^2 + v_2^2)\frac{n}{4} +  v_3^2 \frac{n^2}{4}\\
 &=& \frac{n}{4} + v_3^2
\frac{n^2}{4} \left ( 1 - \frac{1}{n} \right ),\nonumber
\end{eqnarray}
\textit{where $v_{1,2,3}$ are the components of $\bm{v}$.
This leads to Eq. (\ref{tr}) after some simple algebra.}

Therefore, for arbitrary states
\begin{equation}
(\Delta j_\perp )^2 =  \frac{j}{2} + \frac{2j-1}{2j} \int
d^2 \Omega P (\Omega ) \langle j, \Omega |  j_\perp | j, \Omega
\rangle^2 ,
\end{equation}
so that the SU(2) squeezing criterion (\ref{sc}) for $j>1/2$
is equivalent to
\begin{equation}
\int d^2 \Omega P (\Omega ) \langle j, \Omega |  j_\perp | j,
\Omega \rangle^2 < 0 .
\end{equation}
Since $\langle j, \Omega |  j_\perp | j, \Omega \rangle^2$ is
a positive function we get that SU(2) squeezing implies that
$P (\Omega )$ cannot be a classical probability distribution.

\subsection{Classical bounds}

We derive classical upper bounds for the statistics of the
measurement of arbitrary spin observables. This will be further
particularized to the statistics of angular-momentum components
in Sec. III.

\subsubsection{Bounds on the statistics of classical measurements}

For classical measurements the SU(2) $P$ representative of the POVM
element $\Delta_k $ is an ordinary nonnegative function $P_k (\Omega)
\geq 0$ so that for every $\Omega$
\begin{equation}
\label{pPm}
P_k (\Omega) Q(\Omega) \leq P_k (\Omega) Q_{\mathrm{max}},
\end{equation}
where $Q_{\mathrm{max}}$ is the maximum of the $Q$ function of
the measured state (note that $Q(\Omega)$ is always a positive
well behaved function). Applying this to the first equality in
Eq. (\ref{pm}) we get the following upper bound for the statistics
$p_k$ of classical measurements  \cite{nos1}
\begin{equation}
\label{cbsu2a}
p_k \leq \frac{4 \pi}{2j+1} Q_{\mathrm{max}} \mathrm{tr}
\Delta_k  = \langle j, \Omega |  \rho | j, \Omega
\rangle_{\mathrm{max}} \mathrm{tr} \Delta_k  =
\tilde{\mathcal{P}}_k ,
\end{equation}
where for finite-dimensional systems $\mathrm{tr} \Delta_k$ is always
finite. Equation (\ref{cbsu2a}) can be violated if $P_k (\Omega)$ is
more singular than a delta function or takes negative values. In both
cases Eqs. (\ref{pPm}) and (\ref{cbsu2a}) fail to be true. Therefore,
the violation of condition (\ref{cbsu2a}) is a signature of nonclassical
measurement.

\subsubsection{Bounds on the statistics of classical states}

Next we derive an upper bound for the probability of any outcome $k$
that is to be satisfied by all classical states being measured, so
that its violation becomes a sufficient (but not necessary) criterion
of nonclassical behavior concerning the observed state. For classical
states $P (\Omega)$ is an ordinary nonnegative function so that
\begin{equation}
P (\Omega) Q_k (\Omega) \leq P (\Omega) Q_{k,\mathrm{max}},
\end{equation}
where $Q_{k,\mathrm{max}}$ is the maximum of the $Q$ function $Q_k
(\Omega)$ of the POVM element $\Delta_k$. Applying this to the last
equality in Eq. (\ref{pm}) we get the following upper bound for the
probability $p_k$ of the outcome $k$
\begin{equation}
\label{cbsu2b}
p_k \leq \frac{4 \pi}{2j+1} Q_{k,\mathrm{max}} = \langle j, \Omega
| \Delta_k | j, \Omega \rangle_{\mathrm{max}} = \mathcal{P}_k .
\end{equation}
that holds for every $P (\Omega)$ compatible with classical
physics. If this condition is violated for any $k$ the state
is not classical.

\section{Nonclassicality in the measurement of angular-momentum
components}

Next we apply the above approach to the particular case of the
measurement of angular-momentum components. By SU(2) symmetry
we can choose any component without loss of generality, say $j_3$.
In such a case $\Delta_m = |j,m \rangle \langle j,m |$ with
$\mathrm{tr} \Delta_m = 1$ so that the upper bound for classical
measurements is
\begin{equation}
\label{jncm}
p_{j,m} \leq \langle j, \Omega | \rho | j, \Omega
\rangle_\mathrm{max} = \tilde{\mathcal{P}}_{j, \rho},
\end{equation}
where $\rho$ is the state being measured, and the upper bound
for classical states is
\begin{equation}
\label{jncs}
p_{j,m} \leq \left | \langle j, m | j, \Omega \rangle
\right |^2_\mathrm{max} =\mathcal{P}_{j,m} .
\end{equation}
Note that both classical bounds are formally identical. From
now on we consider $m \neq \pm j$, since otherwise $|j, m =
\pm j \rangle$ are SU(2) coherent states and the bound for
classical states is trivial $\mathcal{P}_{j,m} = 1$. On the
other hand, since the states $|j, m = \pm j \rangle$ are
angular-momentum classical they define a classical measurement
and the bound (\ref{jncm}) can never be surpassed.

The maximum of
\begin{equation}
| \langle j,m | j, \Omega \rangle |^2 =  \left ( \begin{array}{c}
2j \cr m+j \end{array} \right )\sin^{2(j-m)} \left ( \frac{\theta}{2}
\right ) \cos^{2(j+m)} \left ( \frac{\theta}{2} \right )
\end{equation}
when $\theta$ is varied is obtained for
\begin{equation}
\tan^2 \frac{\theta}{2} =  \frac{j-m}{j+m} ,
\end{equation}
so that the upperbound for the statistics of classical states is
\begin{equation}
\label{suB}
\mathcal{P}_{j,m} = \left ( \begin{array}{c} 2j
\cr j + m \end{array} \right ) \left ( \frac{j - m}{2j}
\right )^{j-m} \left ( \frac{j+m}{2j} \right )^{j+m} .
\end{equation}

\subsection{The measurement of angular-momentum components is
nonclassical}

In Eq. (\ref{jncm}) let us consider that the measured state is
equal to the measuring state, $\rho = \Delta_m = |j, m \neq \pm j
\rangle \langle j , m \neq \pm j |$, so that the probability is
unity $p_{j,m}=1$. On the other hand, the maximization in
Eq. (\ref{jncm}) is exactly the same we have just carried out so
that the upperbound for the statistics of classical measurements
is
\begin{equation}
\label{suP}
\tilde{\mathcal{P}}_{j,m} = \left ( \begin{array}{c} 2j
\cr j + m \end{array} \right ) \left ( \frac{j - m}{2j}
\right )^{j-m} \left ( \frac{j+m}{2j} \right )^{j+m} .
\end{equation}
The minimum upper bound is obtained for $m=0$ for integer $j$ and
$m = \pm 1/2$ for half integer $j$. These outcomes are the best
candidates to observe nonclassicality. More specifically, for
integer $j$ and $m=0$ we get
\begin{equation}
\label{ob}
\tilde{\mathcal{P}}_{j,m=0} = \frac{(2j)!}{j!^2 2^{2j}} \simeq
\frac{1}{\sqrt{\pi j}} ,
\end{equation}
where the approximation holds for $j \gg 1$. In this case
the upper bound $\tilde{\mathcal{P}}_{j,m=0}$ is clearly below 1,
so that Eqs. (\ref{jncm}) and (\ref{suP}) are infringed and the
measurement is not classical.

As a further example let us consider that the measured state
is a classical state such as the equatorial phase-averaged SU(2)
coherent state
\begin{eqnarray}
\rho &=& \frac{1}{2 \pi} \int_{2\pi} d \phi |j,\theta=\pi/2, \phi
\rangle \langle j, \theta = \pi/2, \phi | \nonumber \\
&=&\frac{1}{2^{2j}}
\sum_{m=-j}^j \pmatrix{2j \cr j+m} |j,m\rangle \langle j,m|,
\end{eqnarray}
where $|j,\theta=\pi/2, \phi \rangle$ are the corresponding
equatorial SU(2) coherent states. In this case $\langle j,
\Omega | \rho | j, \Omega \rangle_\mathrm{max}$ is obtained
for $\theta = \pi/2$ for any $\phi$, so that the classical upper
bound (\ref{suP}) becomes
\begin{equation}
\tilde{\mathcal{P}}_{j,\rho} = \langle j, \Omega | \rho | j, \Omega
\rangle_\mathrm{max} = \frac{1}{2^{4j}} \sum_{m=-j}^j \pmatrix{2j
\cr j+m}^2 ,
\end{equation}
while the statistics is
\begin{equation}
p_{j,m} = \langle j, m | \rho | j, m \rangle = \frac{1}{2^{2j}}
\pmatrix{2j \cr j+m} .
\end{equation}
In Fig. 1 we have represented $p_{j,m}$ (diamonds joined by a
solid line) along with $\tilde{\mathcal{P}}_{j,\rho}$ (dashed
line) for $j=10$, showing that the classical bound is infringed
by the probabilities of the outcomes $m=0,\pm 1$. For example,
for $m=0$ we have $p_{j=10,m=0} = 0.18$ while
$\tilde{\mathcal{P}}_{j=10, \rho} = 0.12$, so that the classical
upper bound is infringed by a 50 \% . As a further example, for
$j=1$ we get $p_{j=1,m=0} = 0.5$, while
$\tilde{\mathcal{P}}_{j=1,\rho} = 0.37$.

\begin{figure}
\begin{center}
\includegraphics[width=8cm]{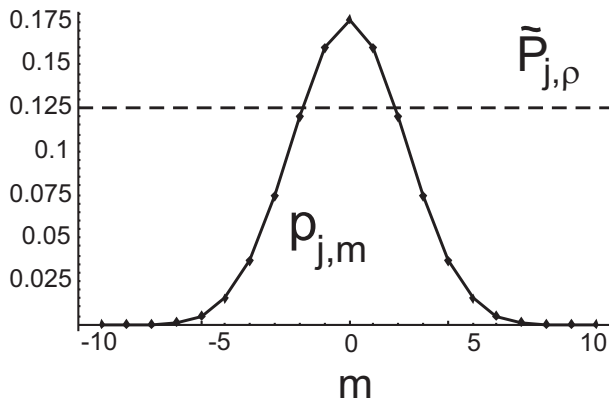}
\end{center}
\caption{Plot of $p_{j,m}$ (diamonds joined by a solid line) and
$\tilde{\mathcal{P}}_{j,\rho}$ (dashed line) for a phase averaged SU(2)
coherent state with $j=10$ showing that the classical bound is
clearly infringed by the statistics of the outcomes $m=0,\pm 1$.}
\end{figure}

\subsection{SU(2) bounds are different from bosonic bounds}

Let us focus on the bounds for classical states via measurement
of an angular-momentum component in Eq. (\ref{suB}). These SU(2)
bounds $\mathcal{P}_{j,m}$ are different from the bounds
$\mathcal{P}^\prime_{j,m}$ for the same statistics derived
from quadrature $P$ and $Q$ functions associated to the
bosonic realization (\ref{So}). This was obtained in Eq. (5.5)
of Ref. \cite{nos1} as
\begin{equation}
\label{qP}
\mathcal{P}_{j,m}^\prime = \frac{( j + m )^{j + m}}{(j + m)!}
\frac{( j - m )^{j - m}}{(j - m)!} \exp ( - 2j).
\end{equation}
To illustrate this difference in Fig. 2 we have represented
$\mathcal{P}_{j,m}$ (diamonds joined by a solid line) and
$\mathcal{P}_{j,m}^\prime$ (stars joined by a dashed line)
for $j = 10$ as functions of $m$. It is shown that the SU(2)
bounds are clearly above the quadrature bounds.

\begin{figure}
\begin{center}
\includegraphics[width=8cm]{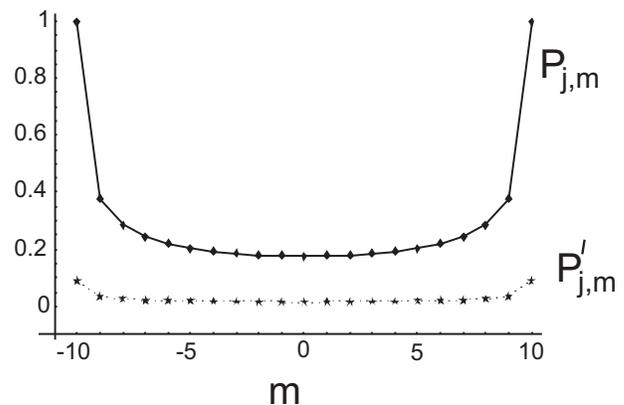}
\end{center}
\caption{Plot of $\mathcal{P}_{j,m}$ (diamonds joined by a solid
line) and $\mathcal{P}_{j,m}^\prime$ (stars joined by a dotted
line) for $j = 10$ as functions of $m$ to illustrate their
difference.}
\end{figure}

The relative difference increases when $j$ increases. This can
be easily seen in the case of integer $j$ and $m=0$ for
example, so that
\begin{equation}
\mathcal{P}_{j,0} = \frac{(2j)!}{2^{2j} j!^2}, \quad
\mathcal{P}^\prime_{j,0} = \frac{j^{2j} \exp(-2j)}{j!^2},
\end{equation}
so that for $j \gg 1$
\begin{equation}
\mathcal{P}_{j,0} \simeq \frac{1}{\sqrt{\pi j}} \gg
\mathcal{P}^\prime_{j,0} \simeq \frac{1}{2 \pi j}.
\end{equation}

These bounds are different because they focus on information
about different variables. As a simple illustrative example
let us consider the case where both the measuring and measured
state are the same SU(2) coherent state $\rho = \Delta_j =
|j,m=j\rangle \langle j ,m=j |$. In this case $p_{j,j} =
\mathcal{P}_{j,j} = 1$ while
\begin{equation}
\mathcal{P}_{j,j}^\prime = \frac{(2j)^{2j}}{(2j)!} \exp (-2j)
\simeq \frac{1}{2 \sqrt{\pi j}},
\end{equation}
where the approximation holds for $j \gg 1$. Therefore the
quadrature bound for classical states $\mathcal{P}_{j,j}^\prime$
is infringed, while the SU(2) bound $\mathcal{P}_{j,j}$ is not.
The state $| j,m=j \rangle$ is clearly not classical concerning
photon number statistics (strongly sub-Poissonian), but this is
classical concerning SU(2) properties, as revealed for example
in two-beam interferometry where these states just reach the
standard quantum limit \cite{SQL}.

\subsection{Independence of SU(2) squeezing and oscillatory
statistics}

Let us present an example of violation of the upper bounds for
classical states without any other typical nonclassical behavior
such as SU(2) squeezing of the orthogonal components $j_\perp$,
nor oscillatory statistics of the measured observable $j_3$. To
this end let us consider the measured state for integer $j>2$
\begin{equation}
| \psi \rangle = \alpha |j,j \rangle + \beta |j,0 \rangle ,
\end{equation}
with $|\alpha |^2 + | \beta |^2 =1$, while the measurement is
$\Delta_0 = | j,0 \rangle \langle j,0|$. The violation of the
upper bound  for classical states (\ref{suB}) holds when
\begin{equation}
\label{nocast}
p_{j,0} = |\beta |^2 > \frac{1}{2^{2j}} \pmatrix{2j
\cr j}.
\end{equation}

\textit{Let us apply to this state the most general SU(2)
squeezing criterion in Eq. (\ref{sc}).} For all $\alpha \neq 0$
the most general $j_\perp$ is of the form
\begin{equation}
j_\perp = \cos \theta j_1 + \sin \theta j_2 .
\end{equation}
\textit{To compute $\langle \psi | j_\perp^2 | \psi \rangle$
let us resort to the bosonic realization (\ref{So}) so that}
\begin{equation}
j_\perp = \frac{1}{2} \left ( a^\dagger_2 a_1 e^{i \theta} +
a^\dagger_1 a_2 e^{-i \theta} \right ) ,
\end{equation}
\textit{and, taking in this case $n=j$ since $j$ is integer,}
\begin{equation}
| \psi \rangle = \alpha |2 n\rangle | 0 \rangle  + \beta
|n \rangle |n \rangle .
\end{equation}
\textit{This allows us to conclude easily that} for all $\theta$
\begin{equation}
\left ( \Delta j_\perp \right )^2 = \frac{1}{2}
\left ( | \beta |^2 j^2 + j \right ) \geq \frac{j}{2},
\end{equation}
so that the weakest squeezing criterion (\ref{sc}) is
never satisfied. Besides, there is no oscillatory statistics
of the measured observable $j_3$ since there are just two
outcomes $m=0,j$.

\subsection{SU(2) Schr\"{o}dinger cat states}

This is the coherent superposition of antipodal SU(2) coherent
states, also known as NOON states \cite{noon}. In the $| j,m
\rangle$ and photon number $|n_1 \rangle |n_2 \rangle$ bases
they can be expressed as
\begin{equation}
\label{Scs}
| \psi \rangle = \frac{1}{\sqrt{2}} \left ( |j, j \rangle
+ |j , -j \rangle \right ) =
\frac{1}{\sqrt{2}} \left ( |n \rangle | 0 \rangle
+ |0 \rangle |n \rangle \right ) ,
\end{equation}
with $j=n/2$. In this case the nonclassical behavior is
revealed by the statistics of $j_1$
\begin{equation}
p_{j,m} = \frac{2}{2^{2j}} \pmatrix{2j \cr j+m}
\end{equation}
for even $j+m$ and $p_{j,m}=0$ otherwise. In Fig. 3 we have
represented $p_{j,m}$ (diamonds joined by solid line) and the
SU(2) bound for classical states $\mathcal{P}_{j,m}$ (stars
joined by a dotted line) for $j=10$. The plot shows that for
$m=0, \pm 2$ there is a clear violation of the classical state
upper bounds. In particular, for $m=0$ we get $p_{j=10,m=0} =
0.35$, while $\mathcal{P}_{j=10,m=0} = 0.18$, so that the
classical upper bound is infringed by a 100 \%.

\begin{figure}[t!]
\begin{center}
\bigskip
\includegraphics[width=8cm]{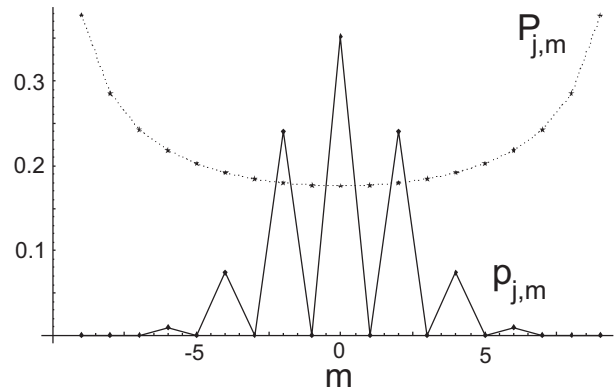}
\end{center}
\caption{Plot of the $j_1$ statistics $p_{j,m}$ (diamonds
joined by solid line) for the Schr\"{o}dinger cat state
(\ref{Scs}) and the SU(2) bound for classical states
$\mathcal{P}_{j,m}$ (stars joined by a dotted line) in
Eq. (\ref{suB}) for $j=10$ showing that for $m=0,\pm 2$
there is a clear violation of the classical upper bound.}
\end{figure}

The nonclassical behavior can be ascribed in this case to the
oscillatory statistics of the measured observable $j_1$ as a
result of the interference of probability amplitudes in the
coherent superposition in Eq. (\ref{Scs}). The interference
minima $p_{j,m}=0$ are compensated by the maxima, where
$p_{j,m}$ takes twice the value for the corresponding SU(2)
coherent state. Thus, the vanishing of $p_{j,m}$ for some $m$
forces the other $p_{j,m}$ to raise above the classical limit.

Concerning SU(2) squeezing we have that $\langle \bm{j}
\rangle = \bm{0}$, so that there is no parallel nor orthogonal
components and the above squeezing criteria fail to be defined.
Anyway, the weakest squeezing criterion (\ref{sc}) is not
satisfied for any component since
\begin{equation}
\left ( \Delta j_3 \right )^2 = j^2, \quad
\left ( \Delta j_{1,2} \right )^2 = j /2 ,
\end{equation}
\textit{as it can be easily computed using the bosonic
realization (\ref{So})}. Nevertheless, these states provide
better interferometric resolution than coherent states of
the same mean number of photons \cite{SQL,noon}.

\subsection{SU(2) intelligent squeezed states}

Let us show  that the intelligent states (\ref{eeis}) satisfying
squeezing criterion (\ref{urc}) violate classical state bounds.
In the basis of eigenstates of $j_{\perp,1}$ the solution of
Eq. (\ref{eeis}) is \cite{BM}
\begin{widetext}
\begin{equation}
\label{BM1}
|j,\eta \rangle = \mathcal{N}\sum_{m=-j}^j
\pmatrix{2j \cr j+m}^{-1/2}  \left [ \frac{4(1-\eta^2 )}{\eta^2}
\right ]^{(j+m)/2} P_{j+m}^{(-m,-m)} \left (
\frac{1}{\sqrt{1-\eta^2}} \right ) |j,m \rangle ,
\end{equation}
\end{widetext}
where $\mathcal{N}$ is a normalization constant and
$P_\ell^{(m,n)} (x)$ are the Jacobi polynomials.

In Fig. 4 we have represented the statistics of $j_{\perp,1}$
(diamonds joined by solid line) for $j=10$, $\eta =0.5$ along
with the SU(2) upper bound for classical states (\ref{suB})
(stars joined by dotted line) showing nonclassical behavior
for $m=0,\pm 1$. In particular for $m=0$ we have $p_{j=10,m=0}
=0.26$ while the classical state bound is $\mathcal{P}_{j=10,m=0}
=0.18$, this is a 44 \% violation of the classical bound.

In Fig. 5 we have plotted the probability $p_{j=10,m=0}$ (solid
line) for the state (\ref{BM1}) as a function of $\eta$ along
with the SU(2) classical state upper bound (\ref{suP})
$\mathcal{P}_{j=10,m=0}$ (dashed line) showing nonclassical
behavior for all $\eta <1$. The state tends to be classical as
$\eta \rightarrow 1$ since in such a case it approaches an SU(2)
coherent state.

\begin{figure}
\begin{center}
\includegraphics[width=8cm]{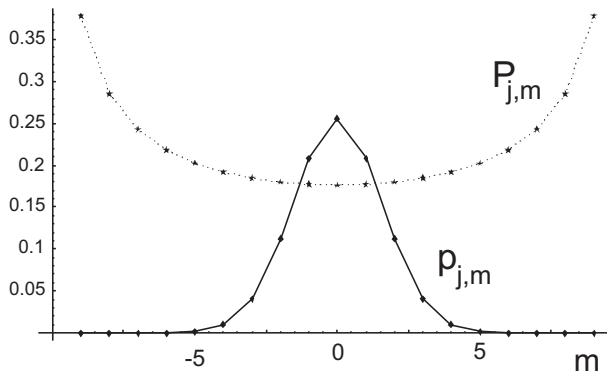}
\end{center}
\caption{Plot of the statistics of $j_{\perp,1}$ (diamonds joined
by solid line) for the state (\ref{BM1}) for $j=10$, $\eta =0.5$
along with the SU(2) classical state upper bound (\ref{suB}) (stars
joined by dotted line) showing the violation of the classical bounds
for small $|m|$.}
\end{figure}

\begin{figure}
\begin{center}
\includegraphics[width=8cm]{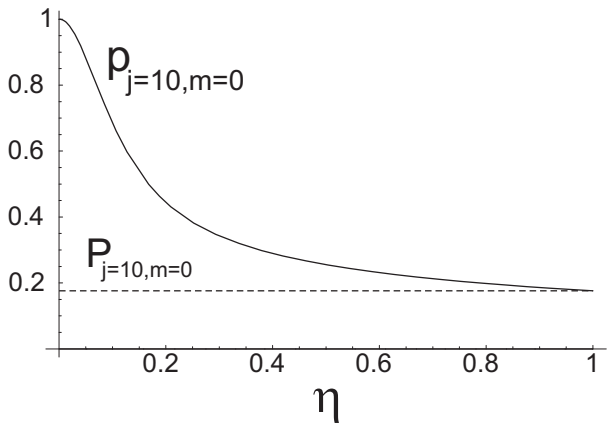}
\end{center}
\caption{Plot of the probability $p_{j,m=0}$ (solid
line) of the eigenvalue $m=0$ of  $j_{\perp,1}$ for the state
(\ref{BM1}) with $j=10$ as a function of $\eta$ along with
the SU(2) classical upper bound (\ref{suP}) $\mathcal{P}_{j,m=0}$
(dashed line) showing nonclassical behavior for all $\eta <1$.}
\end{figure}

\subsection{Bright limit}

Next we derive suitable formulas for the limit of a large
number of photons $j \gg 1$. Besides we focus on the most
favorable cases to violate the classical state upper bounds,
this is  $|m| \ll j$. By using the Stirling approximation
we get the following bright limit for the classical bound
$\mathcal{P}_{j,m}$ in Eq. (\ref{suB})
\begin{equation}
\label{bl}
\mathcal{P}_{j,m} \simeq \sqrt{\frac{j}{\pi (j^2 -m^2)}}
\simeq \frac{1}{\sqrt{\pi j}} .
\end{equation}
For $j \gg 1$ the discrete outcomes $m$ are better described
by a continuous variable $x$, so that $j_1$ for instance
behaves like a single-mode quadrature operator $X$ \cite{k,yk,p2}
\begin{equation}
j_1 \simeq \sqrt{2j} X , \qquad m \simeq \sqrt{2j} x .
\end{equation}
The probability distributions $p_m$ and $p(x)$ are related
in the form
\begin{equation}
\label{blp}
p(x) \simeq \sqrt{2j} p_{j, m =  \sqrt{2j} x} .
\end{equation}
The corresponding classical upperbound for the statistics $p(x)$
derived from (\ref{bl}) and (\ref{blp}) are, respectively
\begin{equation}
p(x) \leq \mathcal{P} = \sqrt{\frac{2}{\pi}} .
\end{equation}
The bound $\mathcal{P}$ coincides with the bound for quadrature
measurements derived from the quadrature $P$, $Q$ functions
\cite{nos1}. This is to say that in this limit angular-momentum
nonclassicality is equivalent to quadrature nonclassicality
\cite{k}.

\section{Conclusions}

We have provided feasible practical procedures to reveal the
nonclassical behavior of angular-momentum states and measurements.
Among other practical situations in quantum optics this includes
two-beam interference and polarization.

A key point is that this approach refers exclusively to the
nonclassical properties of angular momentum, being insensible
to the nonclassical behavior of other variables such as total
photon number. In this regard we have shown that the
nonclassical test derived from SU(2) variables are more
stringent than the one derived from the quadrature $P$, $Q$
function for the same measurement.

The nonclassical tests proposed in this approach are
exceedingly simple since definite conclusions are obtained
without evaluation of moments, or any other more sophisticated
data analysis. They are practical since they refer directly to
the statistics of the measurement. Moreover, we have demonstrated
that these nonclassical tests are independent of other
typical quantum signatures such as SU(2) squeezing or
oscillatory statistics.

\section*{Acknowledgments}

We thank financial support from project QUITEMAD S2009-ESP-1594
of the Consejer\'{\i}a de Educaci\'{o}n de la Comunidad de Madrid.
A. R. acknowledges MICINN FIS2009-10061. A. L. acknowledges support
from project No. FIS2008-01267 of the Spanish Direcci\'{o}n General
de Investigaci\'{o}n del Ministerio de Ciencia e Innovaci\'{o}n.

\end{document}